\documentclass[aps,prl,showpacs,twocolumn,superscriptaddress]{revtex4}
\usepackage{amsmath}
\usepackage[colorlinks]{hyperref}
\usepackage{amssymb}
\usepackage{color}
\usepackage{graphicx,amsmath}
\UseRawInputEncoding \input
\hypersetup{colorlinks,citecolor=red,linkcolor=blue,urlcolor=blue}

\begin{document}

\title{Effect of superconductivity on the shape of flat bands}
\author{V. R. Shaginyan}
\email{vrshag@thd.pnpi.spb.ru} \affiliation{Petersburg Nuclear
Physics Institute of NRC "Kurchatov Institute", Gatchina, 188300,
Russia}\affiliation{Clark Atlanta University, Atlanta, GA 30314,
USA}\author{A. Z. Msezane}\affiliation{Clark Atlanta University,
Atlanta, GA 30314, USA}\author{\fbox{M. Ya.
Amusia}}\affiliation{Racah Institute of Physics, Hebrew University,
Jerusalem 91904, Israel} \affiliation{Ioffe Physical –Technical
Institute, RAS, St. Petersburg 194021, Russia} \author{G. S.
Japaridze}\affiliation{Clark Atlanta University, Atlanta, GA 30314,
USA}

\begin{abstract}
For the first time, basing both on experimental facts and our
theoretical consideration, we show that Fermi systems with flat
bands should be tuned with the superconducting state. Experimental
measurements on magic-angle twisted bilayer graphene of the Fermi
velocity $V_F$ as a function of the temperature $T_c$ of
superconduction phase transition have revealed $V_F\propto
T_c\propto 1/N_s(0)$, where $N_s(0)$ is the density of states at
the Fermi level. We show that the high-$T_c$ compounds $\rm
Bi_2Sr_2CaCu_2O_{8+x}$ exhibit the same behavior. Such observation
is a challenge to theories of high-$T_c$ superconductivity, since
$V_F$ is negatively correlated with $T_c$, for $T_c\propto
1/V_F\propto N_s(0)$. {We show that the theoretical idea of forming
flat bands in strongly correlated Fermi systems can explain this
behavior and other experimental data collected on both $\rm
Bi_2Sr_2CaCu_2O_{8+x}$ and twisted bilayer graphene.} Our findings
place stringent constraints on theories describing the nature of
high-$T_c$ superconductivity and the deformation of flat band by
the superconducting phase transition.
\end{abstract}
 \maketitle

\section{Introduction}

Recent experimental data revealed that in the superconducting state
(SC) the dispersionless flat band vanishes, possessing a finite
Fermi velocity $V_F$,  $V_F\propto T_c$ \cite{mac}. This
observation is in contrast with commonly accepted theoretical
considerations, since $V_F$ is always negatively correlated with
the superconducting phase transition temperature $T_c$ \cite{mac}.
Thus, the experimental fact represents a challenging problem for
the corresponding theory. Such a behavior is typical of Fermi
systems with fermion condensation (FC) \cite{phys_rep}, and was
predicted about twenty years ago \cite{plal}. The reason is that
Fermi systems with flat bands should be tuned with the
superconducting state: The dispersionless flat band vanishes, the
effective mass $M^*$ becomes finite and the corresponding Fermi
velocity $V_F=p_F/M^*\propto T_c\propto \Delta$, with $p_F$ and
$\Delta$ being the Fermi momentum and the superconducting gap,
respectively. Flat bands in Fermi systems with strong interaction
were predicted about thirty years ago \cite{ks}, and the properties
of such systems are predicted and described in details, see e.g.
\cite{vol,univ,prl,phys_rep_94,phys_rep,book,book_20,Khod_2020,shag2020,shag20}.
Topological approach turns out to be a powerful method to gain
information about a wide class of physical systems. Understanding
the topological properties allows one to augment the general
knowledge about physical systems without solving specific
equations. The microscopic approach to a heavy fermion (HF) metal
(for example, computer simulations) gives only particular
information about a specific many-body system, but not about the
universal features, inherent in wide class of HF compounds
\cite{phys_rep,book,book_20}.

HF compounds can be viewed as a new state of matter, since their
behavior near the topological fermion condensation quantum phase
transition (FCQPT) acquire important similarities in their
thermodynamic and transport properties, making HF compounds a
universal state of matter. The concept of FCQPT, forming the recent
experimentally discovered flat bands, originated long ago
\cite{ks}. At first, this idea appeared as a mathematical
curiosity, but now the fermion condensation theory, based on FCQPT,
represents an expanding field with numerous applications
\cite{phys_rep_94,vol,univ,prl,book,book_20,phys_rep,Khod_2020,shag2020,shag20}.
Flat bands have been discovered experimentally in magic angle
twisted bilayer graphene (MATBG). Namely, at a magic twist angle
graphene transforms from a weakly correlated Landau Fermi liquid
(LFL) to a strongly correlated two-dimensional electron system
\cite{graph,Lu,Nick,Choi,don}. The flat band is driven by FCQPT and
accompanied by FC transforming the Fermi surface into the Fermi
volume, as depicted in Fig. \ref{Fig1} (a). This flat band becomes
highly susceptible to reconstruction induced by external and
internal factors such as magnetic field $B$, temperature $T$,
including the setting in of phase transitions like
antiferromagnetic, SC, etc, see e.g.
\cite{phys_rep_94,phys_rep,book,book_20,Khod_2020}. In case of the
flat band, the gap $\Delta$ depends linearly on the pairing
interaction constant $g$, $\Delta\propto g$, see e.g.
\cite{ks,phys_rep_94,phys_rep,book,book_20,Khod_2020,volov11,torma}.

The physics of high-$T_c$ superconductors (HTS), being the
mainstream topic for more than thirty years, still seems to be
elusive, and the mechanism of superconductivity in MATBG is
constantly being debated, see e.g. \cite{sadov}. Based on recent
experimental measurements on MATBG with flat band \cite{mac} and
FCQPT, we arrive at the conclusive statements about the underlying
physics of HTS: {Data collected on very different strongly
correlated Fermi systems as $\rm Bi_2Sr_2CaCu_2O_{8+x}$ and MATBG
show that the Fermi velocity $V_F$ is proportional to the
transition temperature $T_c$, $V_F\propto T_c$ \cite{mac,pan}.}
Taking into account that if pairing is mediated by phonons, or
other bosons that are independent of $V_F$, one has that the
density of states (DOS), $N_s(0)\propto 1/V_F\propto T_c$, see e.g.
\cite{mac,bcs,bard,agd}, and we face a challenging problem to
explain the experimental data. We consider MATBG within the
framework of the model of a homogeneous HF liquid \cite{phys_rep}.
This model avoids the complications associated with the anisotropy
of solids and considers both the thermodynamic properties and the
non Fermi liquid (NFL) behavior by calculating the effective mass
$M^*(T,B)$ as a function of  $T$ and $B$ \cite{phys_rep,book_20}.
The model is applicable in our case, since $V_F\sim 10^4$ m/s is a
hundred or more times smaller than electron velocities of isolated
graphene sheets \cite{mac}. Therefore, the wavelength of an
electron is much larger than the distance between carbon atoms of
the sheet; consequently, we can use the well-known jelly model
\cite{phys_rep}.

In this Letter, {for the first time, basing both on experimental
facts \cite{mac,pan} and our theoretical consideration, we show
that Fermi systems with flat bands should be tuned with the
superconducting state: Flat, or approximately dispersionless, band,
possessing almost infinite effective mass $M^*$, vanishes and the
corresponding Fermi velocity becomes $V_F=p_F/M^*\propto T_c\propto
\Delta$. We demonstrate that such a behavior is of general property
and experimentally observed in the HTS $\rm Bi_2Sr_2CaCu_2O_{8+x}$
and graphene.} This behavior is in contrast to the standard case of
BCS, or to other theories of HTS, where the single particle
electron spectrum does not depend on $\Delta$. We also recall that
the thermodynamic properties of Fermi systems with flat bands
generated by FC strongly depend on external parameters like $T$,
$B$, etc. Our findings place strong constraints on theories
describing the high-$T_c$ superconductivity.

\section{Flat band and superconducting state}

We start with considering a Fermi system with FC at $T=0$. We
employ weak BCS-like interaction with the coupling constant $g$ and
analyze the behavior of both the superconducting gap $\Delta$ and
the superconducting order parameter $\kappa({\bf p})$ as $g \to 0$.
Let us write the usual pair of equations for the Green's functions
$F^+({\bf p},\omega)$ and $G({\bf p},\omega)$ \cite{agd}
\begin{equation}\label{zui2}
F^+=\frac{-g\Xi^*}{(\omega -E({\bf p})+i0)(\omega +E({\bf p})-i0)},
\end{equation}
\begin{equation}\label{zui2a}
G=\frac{u^2({\bf p})}{\omega -E({\bf p})+i0}+\frac{v^2({\bf
p})}{\omega +E({\bf p})-i0},
\end{equation}
where $E^2({\bf p})=\xi^2({\bf p})+\Delta^2$, $\xi({\bf
p})=\varepsilon({\bf p})-\mu$. Here, $\varepsilon({\bf p})$ is the
single particle energy and $\mu$ is the chemical potential. The gap
$\Delta$ and the function $\Xi$ are given by
\begin{equation}\label{zui3}
\Delta=g|\Xi|,\quad i\Xi= \int\int_{-\infty }^{\infty }F^+({\bf
p},\omega)\frac{d\omega d{\bf p} }{(2\pi)^4}.
\end{equation}
Here $v^2({\bf p})=(1-{\xi({\bf p})}/{E({\bf p})})/2,\, v^2({\bf
p})+u^2({\bf p})=1$, and simple algebra gives
\begin{equation}\label{zui6} \xi({\bf p})=\Delta\frac{1-2v^2({\bf p})}
{2\kappa ({\bf p})},
\end{equation}
with $\kappa ({\bf p})=u({\bf p})v({\bf p})$ being the
superconducting order parameter \cite{Bogol}. It is directly seen
from Eq. \eqref{zui6} that $\xi\to0$ as soon as $\Delta\to0$,
provided that $\kappa({\bf p})\neq0$ in some region $p_i<p<p_f$;
thus, the band becomes flat in the region, since $\varepsilon({\bf
p})=\mu$ \cite{plal,phys_rep}. Next, we observe from Eqs.
(\ref{zui3}) and (\ref{zui6}) that
\begin{equation}\label{zui7}
i\Xi=\int_{-\infty }^{\infty }F^+({\bf p},\omega )\frac{d\omega
d{\bf p}}{(2\pi)^4}=i\int\kappa({\bf p})\frac{d{\bf p}}{(2\pi)^3}.
\end{equation}
Clearly seen from Eqs. (\ref{zui3}), (\ref{zui6}) and (\ref{zui7})
is that as $g\to0$ the superconducting gap $\Delta\to0$, while
$\xi=0$ and the dispersion $\varepsilon ({\bf p})$ becomes flat.
Then, $\kappa({\bf p})$ remains finite in some region $p_i\leq
p\leq p_f$, making $\Xi$ finite. Thus, in the state with FC
$\Delta$ can vanish while the order parameters $\kappa({\bf p})$
and $\Xi$ are finite. Taking into account Eqs. (\ref{zui3}) and
(\ref{zui6}) we transform Eqs. (\ref{zui2}) and \eqref{zui2a} as
follows
\begin{equation}\label{zui8}
F^+=-\frac{\kappa({\bf p})}{\omega -E({\bf
p})+i0}+\frac{\kappa({\bf p})}{\omega +E({\bf p})-i0}
\end{equation}
\begin{equation}\label{zui8a}
G=\frac{u^2({\bf p})}{\omega -E({\bf p})+i0}+\frac{v^2({\bf
p})}{\omega +E({\bf p})-i0}.
\end{equation}
Here, at the region $p_i\leq p\leq p_f$, occupied by FC, the
factors $v^2({\bf p})$, $u^2({\bf p})=1-v^2({\bf p})$, $v({\bf
p})u({\bf p})=\kappa({\bf p})\neq0$ are determined by the condition
$\varepsilon({\bf p})=\mu$, while $E({\bf p})\to0$
\cite{ks,phys_rep_94,phys_rep}. From Eqs. (\ref{zui8}) and
\eqref{zui8a} it is seen that in the FC state as $g\to0$, the
equations for  $F^+({\bf p},\omega )$ and $G({\bf p},\omega )$ take
the following form in the FC region
\begin{equation}\label{zui9}
F^+({\bf p},\omega )=-\kappa({\bf p})\left[ \frac{1}{\omega +i0}
-\frac{1}{\omega -i0}\right]
\end{equation}
\begin{equation}\label{zui9a}
G({\bf p},\omega )=\frac{u^2({\bf p})}{\omega +i0}+\frac{v^2({\bf
p})}{\omega -i0}.
\end{equation}
Upon integrating $G({\bf p},\omega)$ over $\omega$ we obtain that
$v^2({\bf p})=n({\bf p})$, where $n({\bf p})$ is the quasiparticles
distribution function. From Eq. (\ref{zui3}) $\Delta$ is seen to be
a linear function of the coupling constant $g$
\cite{ks,volov11,torma}. Since the transition temperature $T_c\sim
\Delta$ tends to zero along with $g\to0$, the order parameter
$\kappa({\bf p})$ of the FC state vanishes at finite temperatures
via the first order phase transition \cite{epl}. {Thus, a flat band
represents a special solution of the BSC equations. In contrast to
BSC-like theories, Eq. \eqref{zui6} defines the dependence of
spectrum $\xi$ on $\Delta$, and, as we will see below, leads to
$V_F\propto T_c$ \cite{mac,phys_rep,plal}.}

\begin{figure}[!ht]
\begin{center}
\vspace*{-0.8cm}
\includegraphics [width=0.47\textwidth] {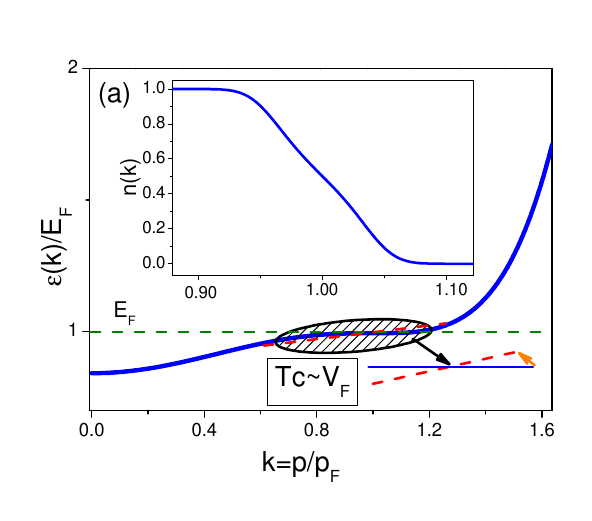}
\includegraphics [width=0.47\textwidth] {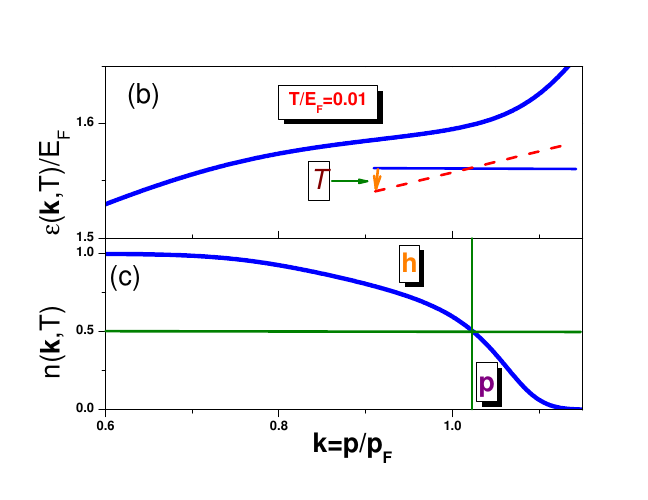}
\end{center}
\vspace*{-0.8cm} \caption{Flat band induced by FC. (a) The flat
single particle spectrum with $V_F=0$ at $T=0$ is shown by the
solid curve. The changed flat band with the emergence of the SC
state with finite $V_F$ is shown by the dashed line, see Eq.
\eqref{zui14}. This change  is depicted by the arrow, and is
schematically shown by the solid and dashed lines. The dashed area
displays the flat band deformed by the superconducting state.
Inset: The quasiparticle occupation number, $n(k)$ at $T=0$ versus
the dimensionless momentum $k=p/p_F$, where $p_F$ is the Fermi
momentum. (b): The single particle spectrum $\varepsilon({\bf
k},T)$ at $T/E_F$; temperature is measured in the units of $E_F$.
(c) The distribution function $n({\bf k},T)$ is asymmetric with
respect to the Fermi level $E_F$, generating specific NFL behavior
and causing the breakdown of both $\mathcal{C}$- and
$\mathcal{T}$-invariance. To clarify the asymmetry, the area
occupied by holes in panel (b) is marked by letter "{\bf h}", while
that for quasiparticles is marked by "{\bf p}".}\label{Fig1}
\end{figure}
At finite $T>0$ the quasiparticle occupation number is given by the
Fermi-Dirac distribution function which is represented in the form
\cite{phys_rep_94}
\begin{equation}\label{FD}
\varepsilon({\bf p},T) -\mu(T)=T\frac{\ln({1-n_0({\bf
p},T)})}{{n_0({\bf p},T)}}.
\end{equation}
Observing that as $T\to 0$, the distribution function satisfies the
inequality $0<n_0({\bf p})<1$ for $p_i\leq p\leq p_f$, we see that
the logarithm is finite and, therefore, the right hand side of Eq.
\eqref{FD} vanishes. Taking into account the well-known Landau
equation $\delta E[n({\bf p})]/\delta n({\bf p})=\varepsilon({\bf
p})$, we conclude  that at $T=0$ Eq. \eqref{zui10} determines
$n_0({\bf p})$ \cite{ks}
\begin{equation}\label{zui10}
\frac{\delta E[n({\bf p})]}{\delta n({\bf p})}=\varepsilon({\bf
p})=\mu;\quad p_i\leq p \leq p_f,
\end{equation}
where $E[n({\bf p})]$ is the Landau functional \cite{agd}. Being
exact \cite{pla} and in accordance with Eq. \eqref{zui6}, Eq.
(\ref{zui10}) describes the state with FC characterized by the
superconducting order parameter $\kappa _0({\bf p})=\sqrt{n_0({\bf
p})(1-n_0({\bf p}))}$ where the functions $n_0({\bf p})$ are
solutions of Eq. (\ref{zui10}). It is instructive to construct
$F^+({\bf p},\omega)$ and $G({\bf p},\omega)$ when $g$ is finite
but small so that the functions $v^2({\bf p})$ and $\kappa({\bf
p})$ can be approximated by the solutions of Eq. (\ref{zui10}). In
that case, $\Xi$, $\Delta$ and $E({\bf p})$ are given by Eqs.
(\ref{zui7}), (\ref{zui3}) and (\ref{zui6}) respectively. Inserting
these into Eqs. (\ref{zui8}) and (\ref{zui8a}), we obtain the
functions $F^+({\bf p},\omega)$ and $G({\bf p},\omega)$.

Let us use Eq. \eqref{zui6} to calculate $M^*$ by
differentiating both sides of this equation with respect to the
momentum $p$ at $p=p_F$. Then
\begin{equation}\label{zui12}
M^*\simeq p_F\frac{p_f-p_i}{2\Delta}.
\end{equation}
When obtaining Eq. \eqref{zui12}, we took into account
that $\kappa(p) = 1/2$ at $p=p_F$, the gap $\Delta(p)$ has a
maximum value $\Delta_1$ at the Fermi surface, and, hence, its
derivative there equals zero. The derivative $d(\nu(p))^2/dp$ was
calculated using the simple estimate $dn(p)/dp\simeq
–1/(p_f-p_i)$.  Since from Eq. \eqref{zui12}
$V_F\propto T_c\propto \Delta$, therefore
\begin{equation}\label{zui14}
V_F\simeq \frac{2\Delta}{p_f-p_i}\propto T_c.
\end{equation}
From Eq. \eqref{zui14} as $T_c\to 0$, the Fermi velocity $V_F\to 0$
and the band becomes exactly flat representing a plateau. When
$T_c$ is finite the plateau is slightly tilted and rounded off at
its end points, as shown in Fig. \ref{Fig1}. At increasing
$\Delta\propto T_c$, both $M^*$ and DOS are reduced and $V_F$
grows. Thus, the electronic system with FC in the superconducting
state is characterized by two effective masses: $M^*$ is the
effective mass given by Eq. \eqref{zui12} and related to the area
$(p_f-p_i)$, with the effective mass $M^*_L$ located at $p\lesssim
p_i$ \cite{phys_rep}. Seen from Fig. \ref{Fig1} (a) is that the
plateau of the flat band of the SC system with FC is slightly
tilted and $M^*$ becomes finite. Indeed, from Eq. \eqref{zui14}
$V_F\propto \Delta\propto T_c$, while at $T\leq T_c$ the Fermi
velocity $V_F$ does not depend on temperature
\cite{book,book_20,phys_rep,plal}. At $T\simeq T_c$ and under the
application of magnetic field, the superconducting phase transition
is of second order, therefore, the singe-particle spectrum is not
changed and $M^*$ does not change at $T\simeq T_c$. Thus, at
$T\simeq T_c$ Eq. \eqref{zui14} is valid, for $M^*$ coincides in
the superconducting state and normal states \cite{mac,phys_rep}. At
$T>T_c$ the slope of the flat band is proportional to $T$, as seen
from Fig. \ref{Fig1} (b). For example, such a dependence can be
measured by using ARPES. Note that the single particle spectrum of
a system with FC is sensitive to the state of the system in
question, and changes with the application of external parameters
like pressure, magnetic field, temperature, etc.
\cite{phys_rep,book,book_20}. It is also seen from Fig. \ref{Fig1}
(b) that both the particle - hole symmetry $\mathcal{C}$ and the
time invariance $\mathcal{T}$ are violated resulting in the
asymmetrical differential tunneling conductivity, and being
suppressed in magnetic fields that drive the system to its LFL
state. This behavior has been predicted and evaluated
\cite{phys_rep,tun,pla_2007}, and turns out to be consistent with
the experimental facts \cite{graph,pan,steg17,pris2008,phys_scr}.

\begin{figure}[!ht]
\begin{center}
\vspace*{-0.8cm}
\includegraphics[width=0.5\textwidth]{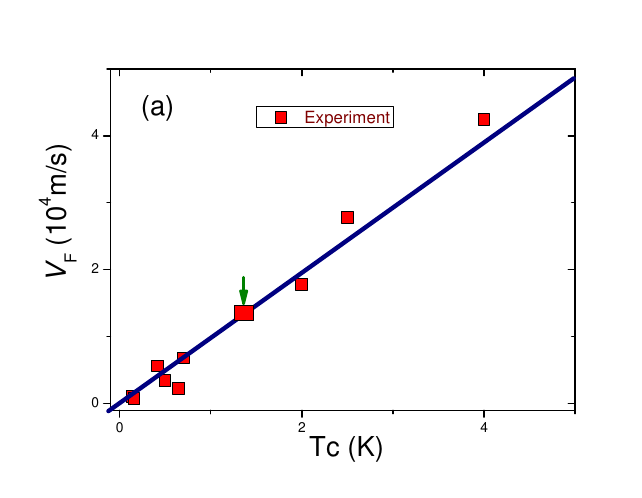}
\includegraphics [width=0.47\textwidth] {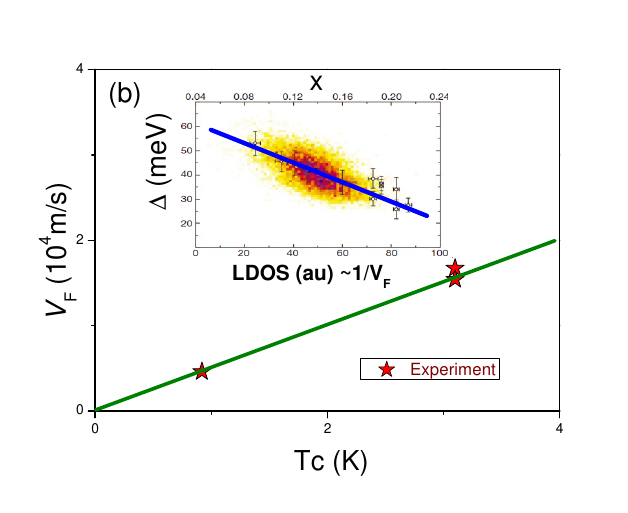}
\end{center}
\vspace*{-0.8cm} \caption{Experimental results for the average
Fermi velocity $V_F$ versus the critical temperature $T_c$ for
MATBG. The results are taken from \cite{mac}. (a) Experiment is
shown by the squares \cite{mac}, the square marked by the arrow
represents experiment \cite{saito}. (b) Experiment is shown by the
stars \cite{mac,yank,step,step20}. Theory is displayed by the solid
line. {Inset is adopted from \cite{pan}, and depicts experimental
dependence of the superconducting gap versus the local density of
states (LDOS) collected on the HTS $\rm Bi_2Sr_2CaCu_2O_{8+x}$;
$\rm x$ is oxygen doping concentration.}} \label{Fig2}
\end{figure}

Measurements of $V_F$ and $N_s(0)$ versus $T_c$ \cite{mac} are
displayed in Figs. \ref{Fig2} (a) and (b) and taken from Refs
\cite{mac,pan,saito,yank,step,step20}. The inset in Fig. \ref{Fig2}
(b) shows experimental facts collected on the HTS $\rm
Bi_2Sr_2CaCu_2O_{8+x}$, with $\rm x$ is oxygen doping concentration
\cite{pan}. The local density of states (LDOS) is shown in
arbitrary units (au), the straight line depicts that LDOS is
inversely proportional to $\Delta$. It is seen from the inset that
the data taken at the position with the highest integrated LDOS has
the smallest gap value $\Delta$ \cite{pan}. These observations in
accordance with Eqs. \eqref{zui12} and Eq. \eqref{zui14}. Thus, our
theoretical prediction  \cite{phys_rep,plal} agrees very well with
the experimental results \cite{mac,pan}. It is noted that $V_F\to
0$ as $T_c\to 0$, seen from Figs. \ref{Fig2}. This result
demonstrates that the flat band is disturbed at finite $\Delta$,
and possesses a finite slope, as seen from Fig. \ref{Fig1} (a).
Clearly, from Figs. \ref{Fig2}, the experimental critical
temperature $T_c$ does not correspond to the Fermi velocity $V_F$
minima as they would in any theory wherein pairing is mediated by
phonons (bosons) that are insensitive to $V_F$ \cite{mac}. Thus,
such behavior is in contrary to that expected within the framework
of the common BSC-like theories that do not assume that the single
particle spectra strongly depend on the pairing correlations
\cite{bard,bcs,mac,phys_rep}. This behavior is based on the
topological FCQPT forming flat bands
\cite{plal,phys_rep,book,book_20}. We note that in case of graphene
a flat band can be disturbed by some external parameters like the
critical angle at which the flat band is emerged. In that case Eq.
\eqref{zui14} is valid at relatively high values of $\Delta$ that
disturb the flat band more strongly than the external parameters
\cite{phys_rep}. $\rm Bi_2Sr_2CaCu_2O_{8+x}$ exhibits this behavior
shown in the inset to Fig. \ref{Fig2} (b).

We now consider  briefly  the entropy $S(T)$ of the system under
consideration, given by the well-known expression \cite{agd}
\begin{equation}
S=-2\int[n_0\ln n_0+(1-n_0)\ln(1-n_0)]\frac{d{\bf
p}}{(2\pi)^3}.\label{ui12}
\end{equation}
The entropy $S$ in Eq. (\ref{ui12}), as the special solution
$n_0({\bf p})$ of Eq. \eqref{zui10}, contains the temperature
independent term $S_0=S(T\to 0)\sim (p_f-p_i)/p_F$. Thus, the
function $n_0({\bf p})$ representing the special solutions of both
BCS and LFL equations determines the NFL behavior. Namely, contrary
to conventional BCS case, the FC solutions are characterized by
infinitesimal value of the superconducting gap, $\Delta \to 0$,
while both $\kappa({\bf p})$ and $\Xi$ remain finite and $S=0$. In
contrast to the usual solutions of the LFL theory, the special
solution $n_0({\bf p})$ is characterized by the entropy $S$
containing the temperature independent term $S_0$. As $T \to 0$
both the normal state of the HF liquid with the finite entropy
$S_{0}$ and the BCS state with $S=0$ coexist being separated by the
first order quantum phase transition represented by the topological
FCQPT, at which the Fermi surface transforms into the Fermi volume,
forming a new quantum liquid \cite{vol}. At this first order phase
transition the entropy undergoes a finite jump $\delta S=S_0$.
Because of the thermodynamic inequality, $\delta Q\leq T\delta S,$
the heat $\delta Q$ of the transition is equal to zero making the
other thermodynamic functions continuous. Thus, at the topological
FCQPT there are no strong critical fluctuations accompanying second
order phase transitions and suppressing the quasiparticles.
Therefore, the quasiparticles survive and define the thermodynamic
and transport properties of HF compounds
\cite{phys_rep_94,phys_rep,book,prl}, rather than quantum critical
fluctuations, see e.g \cite{col}.

\section{Summary}

{The main message of our Letter is that Fermi systems with flat
bands should be tuned with the superconducting state so that Fermi
velocity $V_F=p_F/M^*\propto 1/N_s(0)\propto T_c\propto \Delta$.}
{Such a correlation of the properties of the flat band Fermi
systems with the superconducting state is a new universal effect
and it has been observed recently in graphene \cite{mac} as well as
in the high $T_c$ superconductors $\rm Bi_2Sr_2CaCu_2O_{8+x}$
\cite{pan} that are completely different by their microscopic
structures. The data are well explained in the framework of fermion
condensation theory and to our best knowledge there is no any other
standard theoretical framework that can do the same, since the BSC
like theories state $T_c\propto N_s(0)$ \cite{mac,bard,bcs,agd}.}
Finally, our study of the experimental results \cite{mac,pan}
confirms that the topological FCQPT is the intrinsic feature of
many strongly correlated Fermi systems and can be viewed as the
universal cause of both the NFL behavior and the corresponding new
state of matter \cite{book_20}.

We thank V. A. Khodel for fruitful discussions.
This work was partly supported by U.S. DOE, Division of Chemical
Sciences, Office of Basic Energy Sciences, Office of Energy. This
work is partly supported by the RFBR No. 19-02-00237 À.

\end{document}